\documentclass[12pt,a4paper]{article}
\usepackage{psfig}

\renewcommand\thesection{\arabic{section}.}
\renewcommand\thesubsection{\thesection\arabic{subsection}.}
\renewcommand\thesubsubsection{\thesubsection\arabic{subsubsection}.}
\renewcommand\section[1]{\vspace{\topsep}\vspace{\partopsep}
\refstepcounter{section}
{\par  \noindent\normalsize\bfseries \thesection
\hspace{1em}#1\vspace{\topsep}\par\noindent}}

\newenvironment{refs}
{\vspace{\topsep}\vspace{\partopsep}
{\par \noindent\normalsize\bfseries  References
\vspace{-\topsep}\par\noindent}
\setlength{\parindent}{-5mm}
\begin{list}{}{\topsep 0pt \partopsep 0pt \itemsep 0pt \leftmargin 5mm
\parsep 0pt \itemindent -5mm}}
{\end{list}}

\renewcommand\subsection[1]{
\refstepcounter{subsection}
{\par \protect\vspace{\topsep}\vspace{\partopsep}
 \noindent\normalsize\bfseries \slshape \thesubsection
\hspace{1em}#1\par \noindent}}

\renewcommand\subsubsection[1]{
\refstepcounter{subsubsection}
{\par \protect \vspace{\topsep}\vspace{\partopsep}
\noindent\normalsize \slshape \thesubsubsection
\hspace{1em}#1\par \noindent}}

\newfont{\sansb}{cmssbx10}
\newfont{\sans}{cmss10}
\newcommand{\vol}[2]{$\;\,$\bf #1\rm , #2.} 
\font\fiverm=cmr5

          \font\sixrm=cmr6       
                     
\def\aap{{Astron. Astr.}}                         % DO NOT DELETE
\def\apj{{Astrophys. J.}}                         % DO NOT DELETE
\def\mnras{{MNRAS}}                               % DO NOT DELETE
\def\pasj{{Pub. Astron. Soc. Japan}}              % DO NOT DELETE
\def\teq#1{$\, #1\,$}                           % text equation
{\catcode`\@=11                                                  
\gdef\SchlangeUnter#1#2{\lower2pt\vbox{\baselineskip 0pt\lineskip0pt    
\ialign{$\m@th#1\hfil##\hfil$\crcr#2\crcr\sim\crcr}}}}

\def\Emax{E_{\hbox{\fiverm MAX}}}

\def\dover#1#2{\hbox{${{\displaystyle#1 \vphantom{(} }\over{
   \displaystyle #2 \vphantom{(} }}$}}

\begin{document}

% \begin{flushright}
% contributed paper for Kruger National Park TeV Workshop,\\ 
% to appear in {\it Towards a Major Atmospheric \v{C}erenkov Detector}\\
% ed. O. C. de Jager (Wesprint, Pochefstroom).
% \end{flushright}

\begin{center}
{\large \bf Shock Acceleration and $\gamma$-Ray Emitting Supernova 
            Remnants\vspace{18pt}\\}
  {Matthew G. Baring$^{1,2}$, Donald C. Ellison$^3$, Stephen P. Reynolds$^3$,\\
   Isabelle A. Grenier$^4$ and Philippe Goret$^4$ \vspace{12pt}\\}
{\sl $^1$LHEA, NASA/Goddard Space Flight Center, Greenbelt, MD 20770, USA\\
     $^2$Compton Fellow, Universities Space Research Association\\
     $^3$Department of Physics, North Carolina State University,
        Raleigh, NC 27695, USA\\
     $^4$DSM/DAPNIA/Service d'Astrophysique, CE-Saclay,
        91191 Gif-sur-Yvette, France\\}
\end{center}

\begin{abstract}
Diffusive shock acceleration in the environs of a remnant's expanding
shell is a popular candidate for the origin of SNR gamma-rays, as well
as providing the principal source of galactic cosmic rays.  In this
paper, results from our study of non-linear effects in shock
acceleration theory and their impact on the gamma-ray spectra of SNRs
are presented.  These effects describe the dynamical influence of the
accelerated cosmic rays on the shocked plasma at the same time as
addressing how the non-uniformities in the fluid flow force the
distribution of the cosmic rays to deviate from pure power-laws.  Such
deviations are crucial to gamma-ray spectral determination.  Our
self-consistent Monte Carlo approach to shock acceleration is used to
predict ion and electron distributions that spawn neutral pion decay,
bremsstrahlung and inverse Compton emission components for SNRs.  We
demonstrate how the spatial and temporal limitations imposed by the
expanding SNR shell quench acceleration above critical energies in the
500 GeV - 10 TeV range, thereby spawning gamma-ray spectral cutoffs
that are quite consistent with Whipple's TeV upper limits to the EGRET
unidentified sources that have SNR associations.  We also discuss the
role of electron injection in shocks and its impact on the significance
of electromagnetic components to GeV--TeV spectral formation.
\end{abstract}

\setlength{\parindent}{1cm}

\section{Introduction}
Supernova remnants have long been invoked as a principal source of
galactic cosmic rays, created via the process of diffusive Fermi
acceleration at their expanding shock fronts (e.g.  Drury 1983, Lagage
and Cesarsky 1983).  They can also provide gamma-ray emission via the
interaction of the cosmic ray population with the remnant environment;
this concept was explored recently by Drury, Aharonian and V\"olk
(1994).  In their model, the gamma-ray luminosity is spawned by
collisions between the cosmic rays and nuclei from the ambient SNR
environment.  In the more recent models of Mastichiadis and de Jager
(1996), Gaisser, Protheroe and Stanev (1997), and Sturner, et al.
(1997), $ee$ and $ep$ bremsstrahlung, and inverse Compton scattering
involving shock-accelerated electrons interacting with the cosmic
microwave background and also IR/optical emission (from dust/starlight)
form added components.  See the reviews by de Jager and Baring (1997)
and V\"olk (1997, this volume) for a discussion of various models.

With no definitive detections of gamma-rays from known supernova
remnants, the motivation for modelling these ``hypothetical'' sources
hinges on a handful of spatial associations of unidentified EGRET
sources (Esposito et al. 1996) at moderately low galactic latitudes
with well-studied radio and X-ray SNRs.  These include IC 443, $\gamma$
Cygni and W44, and most have neighbouring dense environments that seem
necessary in order to provide sufficient gamma-ray luminosity to exceed
EGRET's sensitivity threshold.  The remnants associated with several of
the EGRET unidentified sources show an apparent absence of TeV
emission, as determined by Whipple (Lessard et al. 1995, Buckley et al.
1997), which could be explained by an intrinsic cutoff in the
SNR-generated cosmic ray distribution (e.g.  Mastichiadis and de Jager
1996).

All of the above models invoke simple power-law accelerated particle
populations.  In this paper, we utilize the more sophisticated output
of shock acceleration simulations (e.g. Jones and Ellison 1991) to
address the issues of spectral curvature and the maximum energy of
acceleration in the context of SNR gamma-ray emission.  We use output
from the fully non-linear, steady-state Monte Carlo simulations of
Ellison, Baring and Jones (1996) to describe the accelerated particles
in environments where they influence the dynamics of the SNR shell.  Our
results make clear predictions of what maximum energies of gamma-rays
are expected and more accurate predictions of the level of TeV emission
in these sources.  Our model provides a prescription for defining
realistic values of the non-thermal electron/proton abundance ratio.

\section{Fermi Acceleration at SNR Shocks}
Shock acceleration is usually assumed in astrophysical models to
generate power-law particle populations.  This approximation omits the
effect the accelerated particles themselves have on the hydrodynamics
of their shocked environment.  Such non-linear effects, well-documented
in the reviews of Drury (1983) and Jones and Ellison (1991), have a
feedback on the acceleration mechanism and its efficiency.  The slope
of the cosmic ray distributions depends purely on the compression ratio
\teq{r=u_{\rm ups}/u_{\rm down}} of flow speeds either side of the
shock, and this ratio ultimately depends on the shape of the ion
distributions.  Our kinematic Monte Carlo technique (see Ellison,
Baring and Jones 1996) for simulation of Fermi acceleration in the
non-linear regime is ideal for fully exploring particle populations,
acceleration efficiencies and spectral properties.  The simulation
follows particle convection and diffusion in the shock environs using a
simple scattering law: a particle's mean free path \teq{\lambda} is
proportional to its gyroradius \teq{r_g}.  Generalizations of this law
are easily handled by the Monte Carlo simulation.

%    FIGURE 1 GOES HERE
%
\begin{figure}[htb]
\vspace{0.3cm}
\centerline{\psfig{file=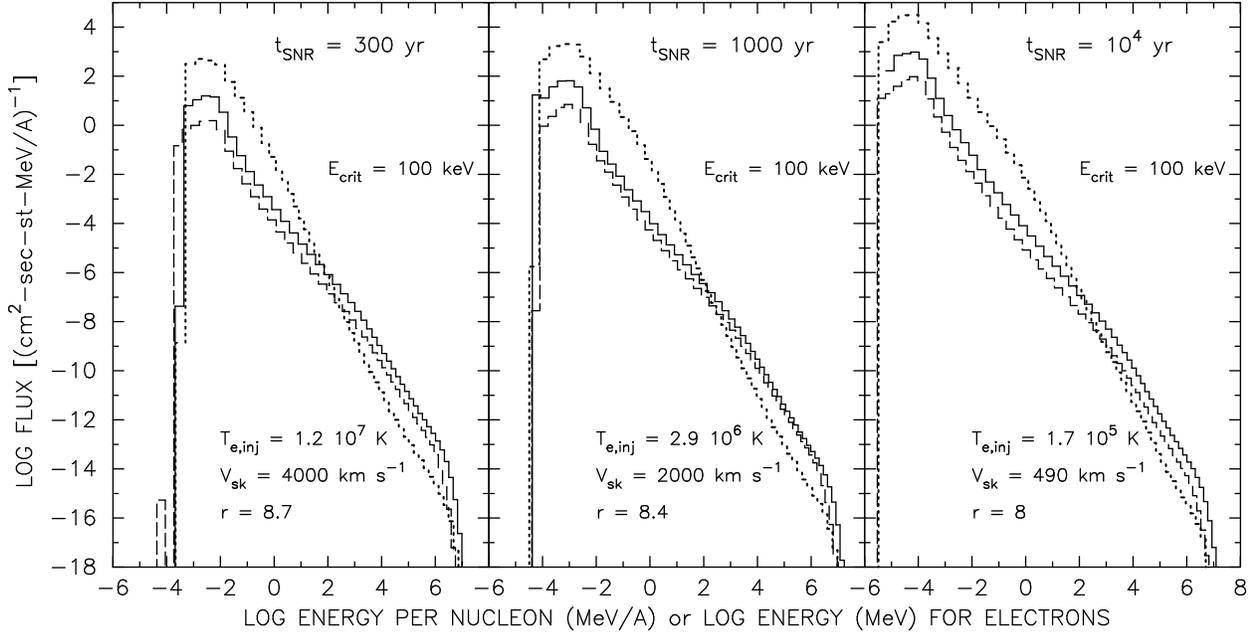,width=16.5cm,angle=270}}
\vspace{-0.3cm}
\caption{Proton, He$^{2+}$ and \teq{e^-} distributions resulting from
the Monte Carlo simulation of acceleration at SNR shocks at three
different ages in the Sedov phase.  The transition to relativistic
energies produces ``bumps'' in the spectra (e.g. at around 1 GeV for
protons).  The ion spectra are the upper two histograms at GeV--TeV
energies (protons -- solid, helium -- dashed), providing nearly all of
the total energy density of the system.  The dashed-dot histograms
represent the $e^-$ distributions (see next section for a discussion of
the parameters \teq{E_{\rm crit}} and the injection temperature
\teq{T_{\rm e,inj}}).  As the shock slows with age, it weakens (\teq{r}
declines).  Upstream parameters were a density of \teq{n_1=1}cm$^{-3}$
and field strength \teq{B_1=3\mu}G.
}
\end{figure}

Clearly the maximum energy of Fermi-accelerated ions is of central
importance to TeV observations.  It can be determined in the Sedov phase by
equating the acceleration time to the remnant age \teq{t_{\hbox{\sixrm
SNR}}}, producing values of the order of (see Baring et al. 1997):
\begin{equation}
E_{\rm max}\; \sim\; 4.7\;\dover{r - 1}{r}\,
   \dover{Q}{\eta}\, \biggl(\dover{B_1}{3\mu {\rm G}}\biggr)\; 
   \biggl( \dover{V_{\rm sk}}{1000\, {\rm km/s}} \biggl)^2\;
   \dover{t_{\hbox{\sixrm SNR}}}{{10^3 {\rm yr}} }\;\, {\rm TeV}
\end{equation}
where \teq{B_1} is the field in units of Gauss, \teq{V_{\rm sk}} is the
expansion speed, and \teq{\eta =\lambda /r_g}.  In the Sedov phase,
where the peak luminosity is expected (see Drury, Aharonian and V\"olk
1994), \teq{\Emax} is actually a slowly increasing function of time
(see Baring et al. 1997).  For young SNRs early in the Sedov phase
(\teq{200}--\teq{5000}yrs), typically \teq{V_{\rm sk}\sim
200}--\teq{3000}km/sec, implying age-limited acceleration terminating
at around \teq{1-10}TeV.  The diffusion scale of the SNR medium is
\teq{\kappa /u} for a diffusion coefficient of \teq{\kappa =\lambda
v/3}, so that Eq.~(1) generally guarantees that maximum diffusion
lengths are always considerably smaller than the shock radius.

Typical simulation output is shown in Fig.~1, where the resulting ion
and $e^-$ distributions are exhibited.  The pressure of the accelerated
ions ($p$ and He$^{2+}$) acts to slow down the fast-moving flow
upstream of the shock, creating a maximum compression ratio $r$ that is
much greater than the canonical ``strong shock'' value of 4.  Such
large $r$ are realized at the largest scales, implying an increased
``efficiency'' of accelerating higher energy particles than those at
lower energies.  Hence upward curvature appears (e.g. see Baring et al.
1997) in the proton, \teq{\alpha} and electron distributions:  high
energy particles have longer mean-free paths and therefore typically
influence the flow on larger scalelengths.  The escape of particles
upstream of the shock (i.e. outside of the remnant's shell), which
occurs on a timescale of \teq{\sim t_{\hbox{\sixrm SNR}}}, gives a
cessation of acceleration, in this case at TeV energies, forcing the
compression ratio to increase to compensate in the flow hydrodynamics.
These non-linear predictions of the spectrum differ significantly from
the test-particle case (Ellison, Baring and Jones 1996), with
deviations from power-law behaviour by as much as a factor of around 3
over four decades in particle energy: this is a significant influence
on predictions of TeV fluxes in gamma-ray SNRs.

\section{Gamma-Ray Emission from Shell-Type Remnants}
In applying the full non-linear Monte Carlo simulation to the
prediction of SNR $\gamma$-ray emission, we model the \teq{\pi^0} decay
component, bremsstrahlung and inverse Compton scattering of background
radiation fields.  Cosmic ray ions collide with nuclei in the cold
ambient ISM to produce \teq{\pi^0}s, which subsequently decay to create
two photons; the decay spectra were calculated much along the lines of
the work of Dermer (1986).  Bremsstrahlung and inverse Compton
contributions were calculated along the standard lines used in Gaisser,
Protheroe and Stanev (1998) and Sturner et al. (1997).  Fig.~2 shows
the emergent $\gamma$-ray spectra resulting from particle distributions
(similar to those in Fig.~1) generated by the non-linear Monte Carlo
simulation.  The non-linear modifications can alter the TeV/EGRET flux
ratio by as much as a factor of 2--3 (Baring, Ellison and Grenier
1997).  In this case, ions are accelerated out to around a few TeV per
nucleon, providing compatibility with the Whipple upper limits.
Turnovers at such low energies are not predicted in the work of
Gaisser, Protheroe and Stanev (1998), but are invoked by Mastichiadis
and de Jager (1996), and Sturner et al. (1997).  One implication of
such low maximum energies for cosmic rays may be that gamma-ray
SNRs are a gamma-ray bright minority of the remnant population, with
other SNRs being required to produce the bulk of cosmic rays out to the
\teq{10^{14}eV} ``knee.''  Validity of such a contention would be 
contingent upon a confirmation that at least some of the EGRET
unidentified sources in Esposito et al. (1996) represent detections
of shell-related emission.  Alternatively, such emission may be common
but at flux levels below EGRET's sensitivity, awaiting future
detections; such a scenario might permit most shell-type $\gamma$-ray
remnants to simultaneously be prolific producers of cosmic rays.

%    FIGURE 2 GOES HERE
%
\begin{figure}[htb]
\vspace{0.0cm}
\centerline{\psfig{file=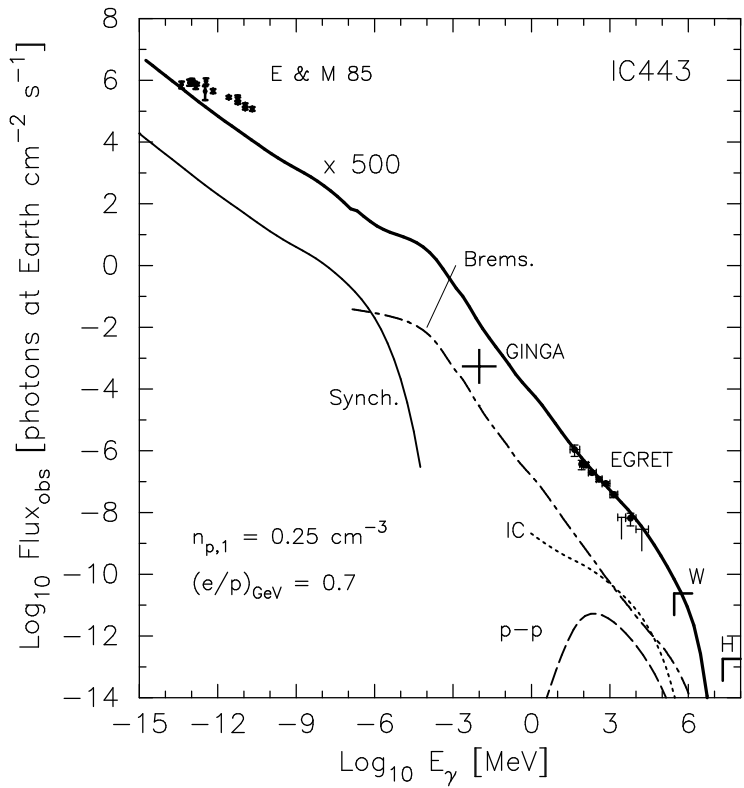,height=10.4cm}}
\vspace{-0.4cm}
\caption{Inverse Compton (IC), bremsstrahlung and pion decay (p-p)
$\gamma$-ray emission (flux) spectral components, and radio to X-ray
synchrotron radiation, obtained by integrating over shock
acceleration-produced ion and electron distributions much like those in
Fig.~1.  These were computed assuming a source volume of 1 pc$^3$
and distance of 1 kpc, and the shock model had parameters
\teq{u_{\rm ups}=876}km/sec, \teq{n_{\rm p,1}=0.25}cm$^{-3}$ and field
strength \teq{B_1=3\mu}G.  Such a model has insufficient density to
compare with the EGRET data for the source associated with IC 443
(Esposito et al. 1996) that are plotted, so we depict via the 
heavy solid curve a scaling of the total spectrum by a factor of 500;
this could be realized through a larger emission volume or source
distance.  Also plotted are radio data from the compilation
of Erickson and Mahoney (1985), the Ginga flux (Wang et al. 1992)
and upper limits from the Whipple (Buckley et al. 1997) and
HEGRA (Prosch et al. 1996) experiments.
}
\end{figure}

Some features of the model results illustrated in Fig.~2 together with
a collection of data for IC 443 that are immediately apparent are that
(i) the low density chosen (following Gaisser, Protheroe and Stanev
1998) inhibits pion decay emission, compared with the high densities
inferred by Drury, Aharonian and V\"olk (1994) in matching EGRET flux
sensitivities, (ii) the strong contribution of the bremsstrahlung in
hard gamma-rays generates a spectrum steep enough to be compatible with
the EGRET source spectral index, and (iii) the 1--10 TeV-range
turnovers in the particle distributions provide total consistency of
the radiation spectrum with Whipple and HEGRA upper limits.  This last
feature refutes recent suggestions that the upper limits obtained by
Whipple to IC 443 are so constraining that they cause serious problems
for the Fermi mechanism (see Baring 1997, and de Jager and Baring 1997,
for more detailed discussions).  Scaling up the source distance and
volume in the figure emphasizes the suitability of this particular
model for the gamma-rays, but provides inconsistencies with the radio
and Ginga X-ray data.  Hence shock acceleration can comfortably
accommodate the gamma-ray data points, and modification of SNR
parameters to reduce the relative importance of bremsstrahlung can
yield compatibility with the X-ray data (discussed in Baring et al.
1997).  The flat radio synchrotron spectral index poses more of a
problem (see Baring et al. 1997).  We note that the localization of
EGRET gamma-rays from IC 443 in Esposito et al. (1996) is entirely
consistent with a shell origin, in contrast to the more confining
localizations available for $\gamma$-Cygni (Brazier et al. 1996,
discussed in de Jager and Baring 1997).  For this reason, we view IC
443 as a better candidate for comparing models with gamma-ray data.

One important aspect of our work is the connection we make between
properties of thermal and suprathermal electrons and the resulting
$e$/$p$ ratio.  We inject electrons as ``test particles'' at
temperature \teq{T_{\rm e,inj}}, which is usually in the keV range (see
Fig.~1) so as (i) to be compatible with temperatures deduced from X-ray
observations, and (ii) to mimic a shock dissipation that deposits a
sizeable fraction of the proton Rankine-Hugoniot thermal energy into
the electron population.  Furthermore, we define a critical energy
\teq{E_{\rm crit}}, above which electrons efficiently resonate with
Alfv\'en-whistler modes (with \teq{\lambda\propto r_g}), and below
which electron diffusion is inefficient, as is expected at suprathermal
energies due to wave damping.  These choices define a range of
possibilities that describe how electrons are injected into the Fermi
process.  Electrons ``injected'' at lower energies (i.e. lower
\teq{T_{\rm e,inj}} and \teq{E_{\rm crit}}) have shorter mean free
paths, and consequently steeper spectra.  Hence these parameters
determine the value of the $e$/$p$ ratio at 1 GeV (and above) which was
around 0.7 for the case exhibited in Fig.~2 (which had \teq{T_{\rm
e,inj}=9.4\times 10^5}K and \teq{E_{\rm crit}=100}keV).  Generally, our
models produce (see Baring et al. 1997) $e$/$p$ ratios considerably
less than unity (e.g. see Fig.~1), which are consistent with cosmic ray
abundances (e.g.  M\"uller et al.  1995) and modelling of the diffuse
$\gamma$-ray background (Hunter et al. 1997), but provide worse fits of
the gamma-ray data for IC 443 that are depicted in Fig.~2.  The shape
of the gamma-ray spectrum below 1 GeV is strongly dependent on the
$e$/$p$ ratio above 1 GeV.  We believe that electron diffusion in
turbulent plasmas favours lower injection energies.  Consequently, we
contend that lower levels (than in Fig.~2) of inverse Compton and
bremsstrahlung emission relative to pion decay radiation are far more
likely (contrasting Sturner et al. 1997) to be representative of
$\gamma$-rays from SNRs.  Notwithstanding, for electrons and protons
accelerated to similar energies above 30 TeV, inverse Compton
scattering is a more efficient creator of super-TeV gamma-rays, and is
expected (Mastichiadis and de Jager 1996) to dominate such signals from
sources like SN1006.  The recent detections of SN1006 by the CANGAROO
experiment above 1.2 TeV, announced at this meeting (see Tanimori et
al., these proceedings), have ushered in a new era for supernova
remnant studies.

\newpage

\section{Conclusion}
In conclusion, the non-linear effects addressed here are desirable
input to any model that invokes Fermi acceleration at SNRs.  They
produce cutoffs in the spectrum that can comfortably accommodate
Whipple's TeV upper limits to unidentified EGRET sources, and provide a
range of realistic estimates of the non-thermal $e$/$p$ ratio.  It is
clear that if some of the EGRET detections turn out to be of gamma-rays
generated in the environs of remnant shells, then gamma-ray emitters
must be a minority of remnants, perhaps mostly young, given that they
cannot produce ions above around a few TeV in profusion.  Remnants that
provide cosmic rays up to the knee must consequently be a gamma-ray
quiet majority.  Alternatively, if fluxes of shell origin are well
below EGRET's and Whipple's flux sensitivities, then the notion that
shell-type remnants are simultaneously gamma-ray bright and prolific
producers of cosmic rays becomes tenable.  It has therefore become
evident that the Whipple and HEGRA upper limits have not destroyed the
hypothesis that shocks in shell-type remnants energize the particles
responsible for the gamma-ray emission, but rather have provided a
powerful tool for constraining our understanding.  Given the recent
detections of SN1006 by CANGAROO, we expect that in the near future,
coupled TeV/sub-GeV, MeV and X-ray observations will discriminate
between the various models, and refine our understanding of gamma-ray
SNRs.

\begin{refs}   %  REFERENCES

\item
Baring, M.~G. 1997, in \it Proc. of the Moriond Workshop on Very High
      Energy Phenomena in the Universe, \rm ed. Tr\^{a}n Thanh V\^{a}n, J.,
      et al. (\'{E}ditions Fronti\`{e}res, Paris), in press.
\item
Baring, M.~G., Ellison, D.~C. \& Grenier, I. 1997, Proc. 2nd Integral Workshop,
   p.~81.
\item
Baring, M.~G., Ellison, D.~C., Reynolds, S.~P., Grenier, I. A.
   \& Goret, P. 1997, to be submitted to \apj
\item
Brazier, K.~T.~S., et al., 1996, \mnras\vol{281}{1033}
\item
Buckley, J.~H. et al. 1997, \aap\ in press.
\item
de Jager, O.~C. \& Baring, M.~G. 1997, in \it Proc. 4th Compton Symposium, \rm 
   ed. Dermer, C.~D. \& Kurfess, J.~D. (AIP Conf. Proc., New York), in press.
\item
Dermer, C.~D.: 1986 \aap\vol{157}{223}
\item
Drury, L.~O'C: 1983 \it Rep. Prog. Phys.\vol{46}{973}
\item
Drury, L.~O'C., Aharonian, F., \& V\"olk, H. 1994, \aap\vol{287}{959}
\item
Ellison, D. C, Baring, M.~G. and Jones, F. 1996, \apj\vol{473}{1029}
\item
Erickson, W.~C., \& Mahoney, M.~J. 1985 \apj\vol{290}{596}
\item
Esposito, J.~A., et al. 1996, \apj \vol{461}{820}
\item
Gaisser, T.~K., Protheroe, R. \& Stanev, T. 1998, \apj\ in press.
\item
Hunter, S.~D., et al. 1997, \apj\vol{481}{205}
\item
Jones, F. C. \& Ellison, D. C. 1991, \it Space Sci. Rev.\vol{58}{259}
\item 
Lagage, P.~O., \& Cesarsky, C.~J. 1983, \aap\vol{125}{249}
\item
Lessard, R.~W., et al.: 1995 \it Proc. 24th ICRC (Rome)\rm , Vol 2, p.~475.
\item
Mastichiadis, A. \& de Jager, O. 1996 \aap\vol{311}{L5}
\item
M\"uller, D., et al.: 1995 \it Proc. 24th ICRC (Rome)\rm , Vol 3, p.~13.
\item
Prosch, C., et al. 1996, \aap\vol{314}{275}
\item
Sturner, S., Skibo, J., Dermer, C.~D., \& Mattox, J.: 1997 \apj\ in press.
\item
Wang, Z.~R., et al. 1992, \pasj\vol{44}{303}
\end{refs}

\end{document}